# Image Reconstruction for Accelerated MR Scan with Faster Fourier Convolutional Neural Networks

Xiaohan Liu, Yanwei Pang, *Senior Member, IEEE*, Xuebin Sun, Yiming Liu,
Yonghong Hou, Zhenchang Wang, and Xuelong Li, *Fellow, IEEE*

*Abstract*—Partial scan is a common approach to accelerate Magnetic Resonance Imaging (MRI) data acquisition in both 2D and 3D settings. However, accurately reconstructing images from partial scan data (i.e., incomplete *k*-space matrices) remains challenging due to lack of an effectively global receptive field in both spatial and *k*-space domains. To address this problem, we propose the following: (1) a novel convolutional operator called Faster Fourier Convolution (FasterFC) to replace the two consecutive convolution operations typically used in convolutional neural networks (e.g., U-Net, ResNet). Based on the spectral convolution theorem in Fourier theory, FasterFC employs alternating kernels of size 1×1 (1×1×1 in 3D case) in different domains to extend the dual-domain receptive field to the global and achieves faster calculation speed than traditional Fast Fourier Convolution (FFC). (2) A 2D accelerated MRI method, FasterFC-End-to-End-VarNet, which uses FasterFC to improve the sensitivity maps and reconstruction quality. (3) A multi-stage 3D accelerated MRI method called FasterFC-based Single-to-group Network (FAS-Net) that utilizes a single-to-group algorithm to guide *k*-space domain reconstruction, followed by FasterFC-based cascaded convolutional neural networks to expand the effective receptive field in the dual-domain. Experimental results on the fastMRI and Stanford MRI Data datasets demonstrate that FasterFC improves the quality of both 2D and 3D reconstruction. Moreover, FAS-Net, as a 3D high-resolution (320×320×256) multi-coil (eight) accelerated MRI method, achieves superior reconstruction performance in both qualitative and quantitative results compared with state-of-the-art 2D and 3D methods.

## I. INTRODUCTION

MAGNETIC Resonance Imaging (MRI) is a non-invasive imaging technique that can provide better soft tissue contrast than many other imaging modalities. Nevertheless, the data acquisition process in MRI is inherently slow due to the sequential measurement collection in the *k*-space domain. This procedure yields a complete *k*-space matrix $\mathbf{K}$ with $\mathbf{K} \in \mathbb{C}^{N \times F \times P}$ for 2D data and $\mathbf{K} \in \mathbb{C}^{N \times F \times P \times S}$ for 3D data where $N$ stands for the number of parallel coil arrays, $F$ is the number of frequency-encoding lines, $P$ and $S$ are respectively the number of phase-encoding lines from different dimensions. Long measurement times reduce patient comfort, cause motion artifacts, and affect patient throughput. Especially in 3D MRI, the acquisition of high-resolution isotropic data will result in an extremely long scanning time, which severely limits the clinical application of related technologies.

A common solution to this problem is to speed up the acquisition process by reducing the number of measurements to obtain an undersampled *k*-space matrix $\overline{\mathbf{K}}$ ( $\overline{\mathbf{K}} \in \mathbb{C}^{N \times F \times P}$ for 2D data, $\overline{\mathbf{K}} \in \mathbb{C}^{N \times F \times P \times S}$ for 3D data). The undersampling process is usually performed along the phase-encoding dimensions. The value of the undersampled position in $\overline{\mathbf{K}}$ is set to zero. Images directly reconstructed from $\overline{\mathbf{K}}$ often have many artifacts due to the non-satisfaction of the Nyquist sampling theorem.

Compressed Sensing (CS) [1] [36] and Parallel Imaging (PI) [23], [24] techniques have been developed to solve the problem of accelerating MRI reconstruction and are available on commercial scanners. Most modern scanners contain multiple receiver coils. In particular, the *k*-space sample measured by the *i*-th coil can be expressed as:

$$\overline{\mathbf{K}}_i = \mathbf{M}\mathcal{F}(\mathbf{S}_i \mathbf{x}) + \boldsymbol{\varepsilon}_i, i = 1, 2, ..., N, \quad (1)$$

where $N$ is the number of coils used for the acquisition, $\boldsymbol{\varepsilon}_i$ is the measurement noise, $\mathcal{F}$ is the Fourier transform operator, $\mathbf{M}$ is the undersampling mask, $\mathbf{S}_i$ is a complex-valued diagonal matrix encoding the position dependent sensitivity map of the *i*-th coil, and $\mathbf{x}$ is the target image to be reconstructed. CS methods usually solve an optimization problem stems from casting the reconstruction as an idealized inverse problem of (1).

PI methods use multi-coil redundant information to interpolate values at unsampled locations in the *k*-space domain (such as GRAPPA [24], RAKI [25], and KerNL [42] *et al.*). The reconstruction weights are estimated using Auto-Calibration Signal (ACS) through an optimization method. However, such methods only exploit local information in a single data, lack the use of different data priors, are overly sensitive to the characteristics of the data itself, and are easily

Submitted in May 2023. This work was supported by the National Natural Science Foundation of China (Grant No. 52227814).

X. Liu, Y. Pang, X. Sun, Y. Liu, and Y. Hou are with the Tianjin Key Laboratory of Brain-Inspired Intelligence Technology, School of Electrical and Information Engineering, Tianjin University, Tianjin 300072, China (e-mails: {lxhlxh,pyw,sunxuebin,yimingliu,houroy}@tju.edu.cn). Y. Hou is also with the Tiandatz Technology Co. Ltd., Tianjin 301723, China.

Z. Wang is with the Beijing Friendship Hospital, Capital Medical University, Beijing 10050, China (e-mails: cjr.wzhch@vip.163.com).

X. Li is with the School of Computer Science and Center for OPTical IMagery Analysis and Learning, Northwestern Polytechnical University, Xi'an, Shaanxi 710072, China (e-mail: li@nwpu.edu.cn).

affected by motion and noise, resulting in a lack of robustness in the reconstruction results. Importantly, PI methods do not work well at high acceleration rates [34].

Recently, more and more methods exploited deep learning [2], [3], [5], [40], [41] for fast MRI reconstruction [16], [17], [18], [19], [20], [21], [22], [35], [38], [39], [43], [44]. Among them, a kind of method uses large-scale datasets to train neural networks to predict *k*-space data [38], [34]. Although *k*-space domain priors for different data are exploited to some extent, all data share the same reconstruction kernel weights, and hence they fail to make full use of the data's respective characteristics. Moreover, the receptive field of the convolutional neural network (CNN, e.g., U-Net [13] and ResNet [14]) is limited, making it challenging to use the information of the ACS when reconstructing the high-frequency region of *k*-space, which further affects the reconstruction performance. We propose a single-to-group *k*-space domain reconstruction algorithm, which can effectively introduce the ACS characteristics of the data itself during deep learning reconstruction and apply it to 3D reconstruction.

MR image reconstruction can be reconstructed in either the image domain or the *k*-space domain, or simultaneously in the two domains. The more mainstream research direction is reconstructing MR images in the image domain or two domains (parallel or sequential) with deep learning. Sriram *et al.* proposed an unrolled iterative reconstruction method to use U-Net for reconstruction in the image domain [7]. Only simple data consistency operations are performed in the *k*-space domain. Complementary information from the *k*-space reconstruction is not used. Eo *et al.* established a cross-domain CNN to sequentially reconstruct MR data in the *k*-space and image domains [8]. It takes advantage of dual-domain reconstruction to a certain extent. However, the algorithm lacks dual-domain interaction and has a limited receptive field. It is hard for frequency domain networks to exploit ACS and symmetric position information. Image domain networks also struggle with global artifacts. Liu *et al.* and Ran *et al.* adopted a parallel reconstruction structure, which can better utilize dual-domain complementary information, but still lacks sufficient receptive fields [9], [10]. To alleviate this problem, we propose a novel convolutional operator called Faster Fourier Convolution (FasterFC) to replace the two consecutive convolution operations common in CNNs (e.g., U-Net, ResNet). FasterFC easily integrates dual-domain reconstruction into one network and dramatically expands the dual-domain receptive field based on the spectral convolution theorem in Fourier theory. In addition, experimental results demonstrate that FasterFC can improve the reconstruction performance and can be used in the sensitivity map estimation network in 2D and 3D tasks.

In addition, we design a split-slice training strategy to cope with the vast computing resource requirements brought by 3D reconstruction. By integrating the single-to-group *k*-space reconstruction algorithm and FasterFC, a new 3D reconstruction framework is proposed. Compared with the traditional Fast Fourier Convolution (FFC) proposed by Chi *et al.* [11], FasterFC has fewer GFLOPs and a much faster calculation speed. Therefore, FasterFC is more desirable for computation-intensive 3D reconstruction tasks.

There are few deep learning based reconstruction methods for 3D fast MRI. One of the primary challenges is the computational burden posed by 3D high-resolution multi-coil isotropic data. The memory usage required for typical 3D multi-coil reconstruction methods can easily surpass the maximum storage capacity of commercially available GPUs. As a result, it is challenging to directly extend the deep learning methods for 2D fast MRI reconstruction to 3D multi-coil tasks.

The novelties and contributions of the paper are as follows:
- A novel convolutional operator called FasterFC, is proposed to replace the two consecutive convolution operations common in CNNs of deep learning based MR image reconstruction from undersampled data. FasterFC results in the global respective field and is much more efficient than the traditional two consecutive convolution operators and classical FFC [11], [12].
- Applying FasterFC in U-Net [13] yields better reconstruction accuracy, training, and inferring efficiency for single-coil fast MRI. Implementing FasterFC in End-to-End Variational Networks (E2E-VarNet, a state-of-the-art multi-coil MR image reconstruction method) [7], consistently improves reconstruction accuracy in terms of NMSE, PSNR, and SSIM without an increase of model size.
- We propose a 3D fast MR image reconstruction method, called FAS-Net, where FasterFC is employed for 3D sensitive map estimation and 3D image reconstruction. Moreover, we propose a single-to-group strategy for 3D reconstruction in *k*-space domain to employ local information in a single volume and group priors in a group of training volumes. The single-to-group *k*-space domain reconstruction together with the FasterFC image domain reconstruction yields promising reconstruction results.

II. PROPOSED METHODS

*A. Faster Fourier Convolution*

FasterFC (Fig. 1) is designed to replace two consecutive convolution operations (Fig. 1(a)) common in CNNs (e.g., U-Net, ResNet). The traditional two consecutive convolution operations have a very small size of the receptive field. By contrast, FasterFC has a global respective field. The 2D and 3D versions of FasterFC are shown in Fig. 1(b) and Fig. 1(c) and are suitable for 2D and 3D reconstruction, respectively.

FasterFC takes an image or spatial domain feature $f_{input}$ as input and outputs a refined image or spatial domain feature $f_{output}$. Firstly, FasterFC employs a 1×1 convolution to map the input $f_{input}$ to a feature map $f_1$ with the number of channels same as the desired output:

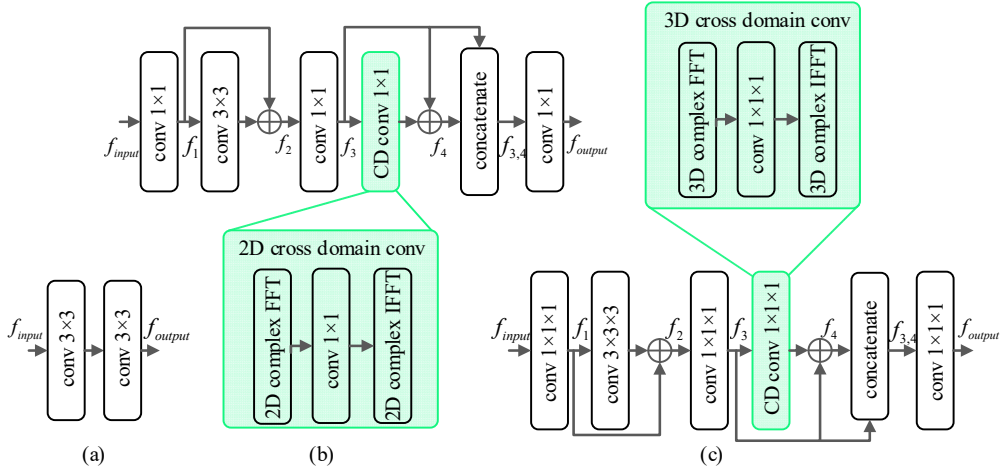

**Fig. 1.** The architecture of (a) traditional two consecutive convolutions, (b) 2D FasterFC, and (c) 3D FasterFC.

$$f_1 = LReLU(InsNorm(Conv_{1\times1}(f_{input})), n), \quad (2)$$

where *LReLU* is leaky ReLU [37], $n$ is the negative slope of the leaky ReLU, and *InsNorm* is instance normalization.

Then $f_1$ is input into a vanilla $3\times3$ convolution with residual connection to expand the local receptive field and refine features in different positions and channels:

$$f_2 = LReLU(InsNorm(Conv_{3\times3}(f_1)), n) + f_1, \quad (3)$$

where $f_2$ is the output of the unique $3\times3$ convolution in FasterFC. Then $f_2$ is processed by a $1\times1$ convolution operation to adjust the channel information and expand the receptive field in the *k*-space domain to all frequencies:

$$f_3 = LReLU(InsNorm(Conv_{1\times1}(f_2)), n), \quad (4)$$

where $f_3$ is the output of the second $1\times1$ convolution in FasterFC. Subsequently, a cross-domain $1\times1$ convolution with residual connection takes $f_3$ as input to extract features in the Fourier domain so that the receptive field of spatial domain input can be expanded into all positions. Generally, we designate the first half of the channels as real channels and the second half as imaginary channels. So $f_3$ can be transformed into the frequency domain using the complex FFT.

After FFT, a frequency domain feature map is obtained, which has the same resolution as $f_3$. The $1\times1$ convolution is then adopted to refine the frequency domain features. Finally, the Inverse FFT (IFFT) transforms the output of the convolution into the spatial domain:

$$f_4 = \mathcal{F}^{-1}(LReLU(InsNorm(Conv_{1\times1}(\mathcal{F}(f_3))), n)) + f_3. \quad (5)$$

$f_3$ and $f_4$ are concatenated together in the channel dimension and the result is used as the input of the last $1\times1$ convolution:

$$f_{3,4} = Concatenate(f_3, f_4). \quad (6)$$

Finally, the last $1\times1$ convolution is used to pick the most suitable features and restore channel dimensions to the predetermined size (half of the input of the last $1\times1$ convolution):

$$f_{output} = LReLU(InsNorm(Conv_{1\times1}(f_{3,4})), n). \quad (7)$$

Traditional two consecutive convolution operations (two $3\times3$ convolutions) have small as $5\times5$ receptive field size. Because FFT and IFFT are performed in FasterFC, FasterFC has global respective field size. As a result of substituting two convolutional layers with a solitary FasterFC, which solely necessitates a single domain transformation, the computational speed of FasterFC is nearly twofold in comparison to the FFC requiring two domain transformations. Notwithstanding, FasterFC maintains its proficiency in feature extraction and reconstruction performance without experiencing any degradation.

*B. Faster Fourier Convolution Based U-Net*

This section details how 2D FasterFC works in U-Net [13] (a common CNN widely used in fast MRI tasks [7], [9]).

U-Net has a symmetrical encoder-decoder structure, where the encoder undergoes multiple downsampling and the decoder employs multi-level upsampling to restore features to their original resolution. Each stage uses two convolution operations for feature extraction in both the encoder and the decoder. FasterFC replaces two consecutive convolution operations in every level of U-Net. This process is convenient and does not change the size of the input and output feature maps for each stage. The resultant network is called FasterFC-U-Net. The structure of U-Net, FasterFC-U-Net, and the replacement process are shown in Fig. 2.

The parameter number of an *L*-level FasterFC-U-Net (downsampling *L* times) is as follows:

$$N_{FastFCu} = 4\times c + 4\times c\times c + (c\times c)\times(3\times3)$$
$$+ \sum_{i=1}^{L} 2^{i-1}c \times 2^i c + 4\times 2^i c \times 2^i c + (2^i c \times 2^i c)\times(3\times3)$$
$$+ \sum_{i=1}^{L} (2^i c \times 2^{i-1}c + 4\times 2^{i-1}c \times 2^{i-1}c \quad (8)$$
$$+ (2^{i-1}c \times 2^{i-1}c)\times(3\times3)) + (\sum_{i=1}^{L} 2^i c \times 2^{i-1}c)\times(2\times2).$$

For an *L*-level U-Net, the number of parameters is $N_u$:

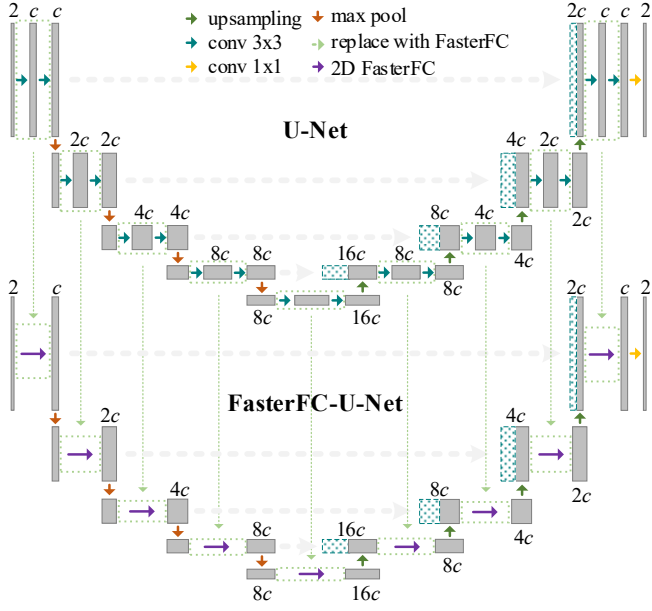

**Fig. 2.** The architecture of U-Net, FasterFC-U-Net, and how FasterFC replaces two consecutive convolution operations in U-Net.

$$N_u = (2 \times c + c \times c + \sum_{i=1}^{L-1} 2^{i-1}c \times 2^i c + 2^i c \times 2^i c) \times (3 \times 3)$$
$$+ 2 \times c + (\sum_{i=1}^{L} 2^i c \times 2^{i-1}c + 2^{i-1}c \times 2^{i-1}c) \times (3 \times 3) \quad (9)$$
$$+ (2^{L-1}c \times 2^L c + 2^L c \times 2^L c) \times (3 \times 3)$$
$$+ (\sum_{i=1}^{L} 2^i c \times 2^{i-1}c) \times (2 \times 2).$$

The ratio $r = N_{FastFCu} / N_u$ between $N_{FastFCu}$ and $N_u$ is a function of $c$ and $L$. The curves of $r$ versus $c$ and $L$ are shown in Fig. 3. For the typical case of $c = 32$ and $L = 4$, $r = 0.86$ holds. That is, the model size of FasterFC-U-Net is 86% of that of U-Net.

Therefore, compared with U-Net, FasterFC-U-Net is more lightweight but has better reconstruction accuracy due to its much larger receptive fields in both the image domain and frequency domain.

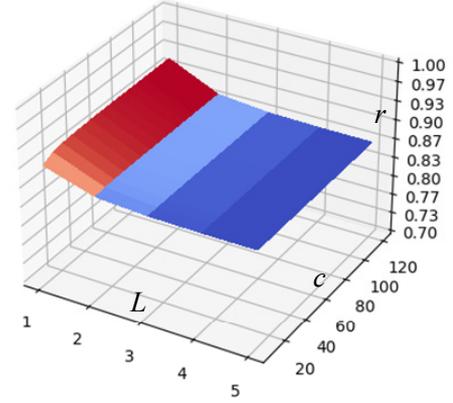

**Fig. 3.** The ratio $r = N_{FastFCu} / N_u$.

### C. Faster Fourier Convolution Based E2E-VarNet

FasterFC can also be used for 2D multi-coil MR image reconstruction. E2E-VarNet [7] is a classic 2D multi-coil MR reconstruction structure. It consists of two parts: one part uses U-Net to predict the sensitivity maps; the other part uses the estimated sensitivity maps and cascaded U-Nets to optimize the MR data iteratively:

$$\mathbf{K}^{t+1} = \mathbf{K}^t - \eta^t \mathbf{M}(\mathbf{K}^t - \bar{\mathbf{K}}^t) + G(\mathbf{K}^t), \quad (10)$$

where $\eta^t$ is used to weight the data based on the distance from the raw measured value to balance the consistency and smoothness of the data. The refinement module $G$ is defined as follows:

$$G(\mathbf{K}^t) = \mathcal{F} \circ \mathcal{E} \circ UNet(\mathcal{R} \circ \mathcal{F}^{-1}(\mathbf{K}^t)), \quad (11)$$

the operator $\mathcal{E}$ denotes the expansion operator, which accepts the input of the image $\mathbf{x}$ and sensitivity maps. In the idealized noise-free scenario, this operator calculates the image perceived by each coil:

$$\mathcal{E}(\mathbf{x}) = (\mathbf{x}_1, ..., \mathbf{x}_N) = (\mathbf{S}_1 \mathbf{x}, ..., \mathbf{S}_N \mathbf{x}). \quad (12)$$

$\mathcal{R}$ is the reduce operator to combine the images from all the coils:

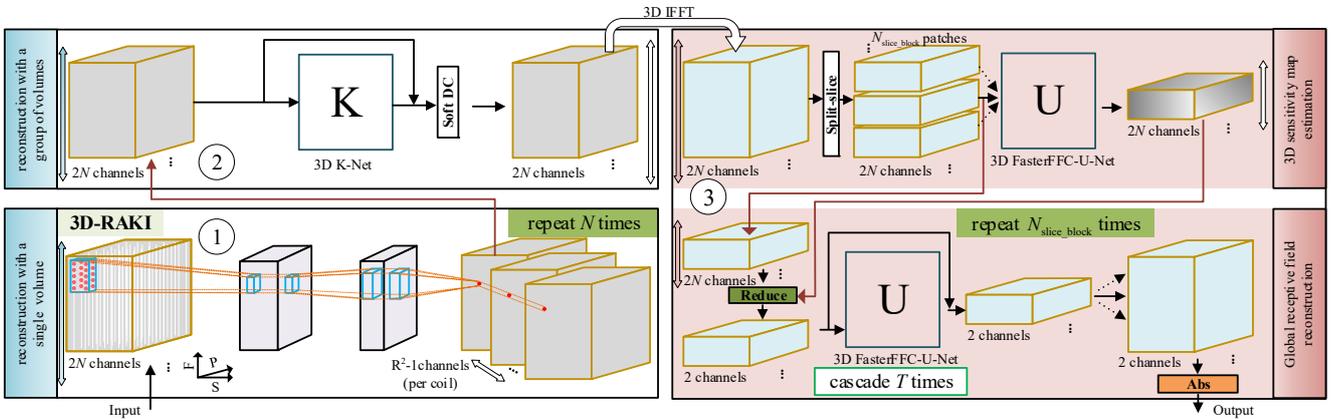

(a) Reconstruction in *k*-space domain with single-to-group algorithm   (b) Reconstruction in image domain with 3D FasterFC-U-Net

**Fig. 4.** The architecture of FAS-Net.

$$\mathcal{R}(\mathbf{x}_1,...,\mathbf{x}_N) = \sum_{i=1}^{N} \mathbf{S}_i^* \mathbf{x}_i, \quad (13)$$

where $\mathbf{S}_i$ are the sensitivity maps estimated by a U-Net, $\mathbf{S}_i^*$ refers to the adjoint matrix of matrix $\mathbf{S}_i$:

$$\mathbf{S} = dSS \circ UNet(\mathcal{F}^{-1} \circ \mathbf{M}_{center}(\bar{\mathbf{K}})). \quad (14)$$

The $\mathbf{M}_{center}$ operator zeros out all lines except for the ACS lines. The $dSS$ operator normalizes the estimated sensitivity maps to satisfy:

$$\sum_{i=1}^{N} \mathbf{S}_i^* \mathbf{S}_i = 1. \quad (15)$$

We propose to modify E2E-VarNet by using FasterFC. The modified method is called FasterFC-E2E-VarNet. Specifically, all U-Nets used in E2E-VarNet are replaced by FasterFC. The U-Net used for sensitivity maps estimation in (14) is replaced by FasterFC-U-Net and the sensitivity map is expressed as:

$$\mathbf{S} = dSS \circ FasterFCUNet(\mathcal{F}^{-1} \circ \mathbf{M}_{center}(\bar{\mathbf{K}})). \quad (16)$$

The U-Nets used for MR image refinement in (11) are replaced by FasterFC-U-Net to achieve dual-domain global receptive field:

$$G(\mathbf{K}^t) = \mathcal{F} \circ \mathcal{E} \circ FasterFCUNet(\mathcal{R} \circ \mathcal{F}^{-1}(\mathbf{K}^t)). \quad (17)$$

Both replacements can effectively improve the reconstruction performance of E2E-VarNet.

*D. FasterFC Based Single-to-group Network for 3D Reconstruction*

We propose to adopt FasterFC for 3D high-resolution multi-coil fast MRI reconstruction. The proposed method is called FAS-Net. The FAS-Net (Fig. 4) begins with reconstruction in *k*-space domain by a single-to-group algorithm (Fig. 4(a)) followed by reconstruction in image domain by FasterFC based 3D sensitivity map estimation and image reconstruction (Fig. 4(b)).

The single-to-group algorithm for *k*-space reconstruction consists of a single reconstruction stage (Fig. 4①) and a group reconstruction stage (Fig. 4②). In the single reconstruction stage, only a single volume (a set of slices) is used for training the learnable parameters independently. In the group reconstruction stage, all the volumes of the training dataset are used for learning parameters so that priors of the whole dataset can be exploited.

Generally, a 3D image reconstruction algorithm is much more time-consuming and memory-consuming than a 2D image reconstruction algorithm. We propose three strategies to tackle these problems. (1) The single reconstruction stage, group reconstruction stage, and FasterFC stage are performed in a separate manner instead of an end-to-end manner. After training the current stage, its model is fixed and its output forms a new dataset for the subsequent stage. (2) Hight-weight networks (3D-RAKI and 3D K-Net) are designed for *k*-space domain reconstruction (i.e., single reconstruction followed by group reconstruction). (3) A split-slice strategy is proposed for image domain reconstruction (i.e., FasterFC stage).

*1) Single Reconstruction Stage of k-space Domain Reconstruction*

Single reconstruction in *k*-space domain is a process of learning volume-specific parameters and reconstructing the volume with the parameters. The core is to embed the local *k*-space prior of an ACS region into regions beyond the ACS region. Classical algorithms such as GRAPPA [24] and RAKI [25] can be used for single reconstruction. But they are oriented to 2D reconstruction. We extend RAKI to a 3D version called 3D-RAKI.

The inference process of 3D-RAKI is shown in the bottom of Fig. 4(a). The input of 3D-RAKI is the undersampled zero-filled *k*-space matrix $\bar{\mathbf{K}} \in \mathbb{R}^{F \times P \times S \times 2N}$. Assume that the accelerate rates of the two orthogonal phase-encoding directions are identical. Let $R$ be the acceleration rate in one direction. The total acceleration rate simultaneous in two directions is $R^2$. The output of the 3D-RAKI for coil $i$ is written as $\tilde{\mathbf{K}} \in \mathbb{R}^{F \times P \times S \times 2}$. There is a 3D CNN model $g_i$ for coil $i$ to map $\bar{\mathbf{K}}$ to $\tilde{\mathbf{K}}$:

$$\tilde{\mathbf{K}}_i = g_i(\bar{\mathbf{K}}). \quad (18)$$

$g_1,...,g_N$ have the same shallow architecture of 3D CNN but their parameters are not shared.

The 3D CNN model for a coil consists of three convolutional layers. The filters of three convolutional layers are expressed as $\mathbf{w}_1, \mathbf{w}_2, \mathbf{w}_3$. The size of $\mathbf{w}_1, \mathbf{w}_2, \mathbf{w}_3$ are $b_1^F \times b_1^P \times b_1^S \times 2N \times n_1$, $b_2^F \times b_2^P \times b_2^S \times n_1 \times n_2$, and $b_3^F \times b_3^P \times b_3^S \times n_2 \times n_{out}$, respectively. Nonlinear activation function ReLU are applied for the first two layers and a linear activation function is applied for the third layer.

The first layer is computed by:

$$F_1(\bar{\mathbf{K}}) = \text{ReLU}(\mathbf{w}_1 * \bar{\mathbf{K}}). \quad (19)$$

The second layer takes the output of the first layer as input and can be computed by:

$$F_2(F_1(\bar{\mathbf{K}})) = \text{ReLU}(\mathbf{w}_2 * F_1(\bar{\mathbf{K}})). \quad (20)$$

With a linear activation function, the output of the third layer can be calculated by:

$$F_3(F_2(F_1(\bar{\mathbf{K}}))) = \mathbf{w}_3 * F_2(F_1(\bar{\mathbf{K}})). \quad (21)$$

It is noted that all the layers employ dilation convolution with dilation rate being $R$ for the purpose of making use of acquired encoding lines.

The network parameters $\mathbf{\theta}_j = \{\mathbf{w}_1, \mathbf{w}_2, \mathbf{w}_3\}$ is trained by gradient descent with backpropagation [28] and momentum [29]. Supervision is applied on the ACS region of $\tilde{\mathbf{K}}$.

By filling all the undersampled *k*-space data of the dataset with 3D-RAKI separately, we obtain a new *k*-space dataset $\{\tilde{\mathbf{K}}_1^{3D},...,\tilde{\mathbf{K}}_I^{3D}\}$ that is initially filled with local priors. Next, group reconstruction can be used for further *k*-space data refinement.

*2) Group Reconstruction Stage of k-space Domain Reconstruction*

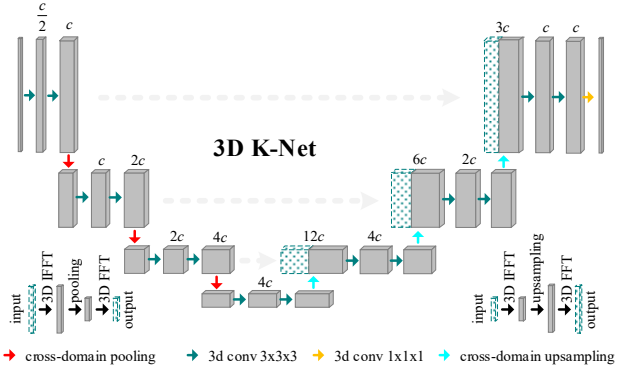

**Fig. 5.** The architecture of 3D K-Net.

To make use of the priors contained in a group of training volumes, we propose to generalize our previous work of K-Net [9]. For learning global priors present in different data, we have extended the K-Net to a 3D version in group reconstruction. A 3D K-Net, featuring a residual connection, is employed to reconstruct $k$-space domain data generated by the 3D-RAKI. The process of group reconstruction is illustrated in Fig. 4②.

The architecture of the 3D K-Net is depicted in Fig. 5, which is nearly identical to that of a traditional 3D U-Net. The primary differences lie in the pooling and upsampling operators, which are replaced by 3D cross-domain pooling (bottom-left of Fig. 5) and 3D cross-domain upsampling (bottom-right of Fig. 5). In the proposed 3D cross-domain pooling, the input is transformed to image domain by 3D IFFT. Pooling is then applied in the image domain. Finally, 3D FFT is employed re-transform the pooled data to frequency domain. The 3D cross-domain upsampling has the same process with the pooling operation instead of upsampling.

In the inference phase, the 3D K-Net with residual connection takes the 3D $k$-space data filled with 3D-RAKI as input, of size $2N \times F \times P \times S$. Then Soft DC [7] criterion is adopted in group reconstruction:

$$k_{rec}(j) = \begin{cases} \bar{k}(j) & \text{if } j \notin \Omega, \\ \bar{k}(j) - \gamma(\bar{k}(j) - k(j)) & \text{if } j \in \Omega. \end{cases} \quad (22)$$

where $j$ represents the index of the vectorized representation of $k$-space data, $\bar{k}$ denotes the output of a 3D K-Net, $k$ means the raw incomplete $k$-space data, $\Omega$ is the index set of sampling data, and $\gamma$ is a trainable hyperparameter to balance predicted data and original sampling data. As defined in (22), if the position of the input $k$-space data point is not sampled ($j \notin \Omega$), the predicted values are used directly. For the sampled points, $\gamma$ is used to weigh the data based on the distance from the original measured value to balance the consistency and smoothness of the data. Finally, 3D IFFT transforms multi-coil refined $k$-space data into the spatial domain and is used as the input to FasterFC-based image reconstruction. This process can be formulated as:

$$\begin{aligned}\bar{\mathbf{x}}^{KNet} &= \mathcal{F}^{-1} \circ \bar{\mathbf{K}}^{KNet} \\ &= \mathcal{F}^{-1} \circ SoftDC \circ (KNet(\tilde{\mathbf{K}}^{3D}) + \tilde{\mathbf{K}}^{3D}).\end{aligned} \quad (23)$$

In the training phase, the output of the 3D K-Net is supervised by the ground truth (denoted by $\mathbf{x}$) of fully sampled images obtained by the complete $k$-space matrices. The following Structural Similarity (SSIM) [4] loss function

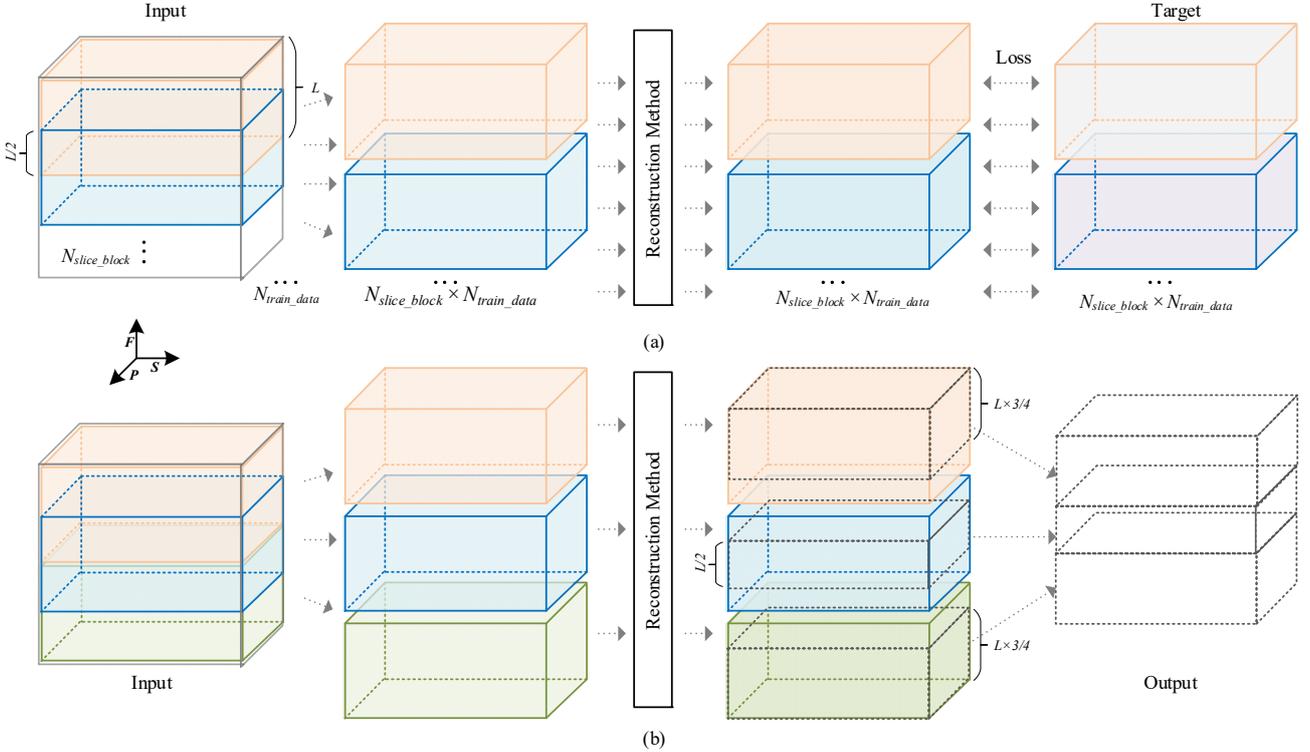

**Fig. 6.** The process of (a) split-slice training and (b) split-slice reconstruction.

$J(\tilde{\mathbf{x}}^{KNet}, \mathbf{x})$ is used to train the 3D K-Net:

$$J(\tilde{\mathbf{x}}^{KNet}, \mathbf{x}) = 1 - SSIM(\tilde{\mathbf{x}}^{KNet}, \mathbf{x}), \quad (24)$$

where $SSIM(\tilde{\mathbf{x}}^{KNet}, \mathbf{x})$ is the value of SSIM between $\tilde{\mathbf{x}}^{KNet}$ and $\mathbf{x}$, $\tilde{\mathbf{x}}^{KNet} = RSS \circ abs \circ \overline{\mathbf{x}}^{KNet}$, and $abs$ computes the absolute value of a complex valued data followed by a root-sum-squares ($RSS$) reduction for each pixel:

$$\tilde{\mathbf{x}}^{KNet} = RSS(\overline{\mathbf{x}}_1^{KNet}, ..., \overline{\mathbf{x}}_N^{KNet}) = \sqrt{\sum_{i=1}^{N} \left|\overline{\mathbf{x}}_i^{KNet}\right|^2}. \quad (25)$$

*3) 3D-FasterFC Based Image Reconstruction*

The output of the K-Net based group reconstruction is used as the input to 3D-FasterFC based image reconstruction. It is composed of 3D-FasterFC based 3D sensitivity estimation (top of Fig. 4(b)) and 3D-FasterFC based image reconstruction (bottom of Fig. 4(b)).

**Split-slice strategy**: Directly applying the proposed FasterFC based method on 3D data is so memory-consuming that the memory consumption can easily exceed the maximum of memory of available GPUs. In fact, the memory consumption problem severely limits the research on 3D reconstruction. To tackle the problem, we propose a split-slice strategy for extract features with 3D convolutional filters. The specific process of the method is demonstrated in Fig. 6.

In the training phase, every $\overline{\mathbf{x}}^{KNet}$ in the train set is split into several small slice blocks. The resolution of the slice block is $N \times L \times P \times S \times 2$ (2 is the complex dimension consisting of the real and imaginary parts). The first slice block starts from the first slice. The next slice block starts from the $L/2$-th slice. That is, the distance of starting position of adjacent slice blocks is $L/2$, and adjacent blocks overlap with $L/2$ slices. So, each $\overline{\mathbf{x}}^{KNet}$ can be split into $2 \times F/L - 1$ blocks, denoted by $N_{\text{slice\_block}}$. For an MR dataset that has $N_{\text{train\_data}}$ high-resolution multi-coil 3D $k$-space data, $N_{\text{slice\_block}} \times N_{\text{train\_data}}$ training data are obtained for split-slice training. The target data $\mathbf{x}$ are processed by the same split-slice operation to get $N_{\text{slice\_block}} \times N_{\text{train\_data}}$ target data. These new target data have a corresponding relationship with the input data, which are used as the ground truth for training the FasterFC-based network. The details of the training process are shown in Fig. 6(a).

In the inference phase described in Fig. 6(b), the input data with size $N \times F \times P \times S \times 2$ is first split into $N_{\text{slice\_block}} = 2 \times F/L - 1$ slice blocks through the split-slice operation. Each slice block with size $N \times L \times P \times S \times 2$ is fed into the reconstruction network for forward propagation, resulting in a reconstructed 3D slice block. Then, all the $2 \times F/L - 1$ refined 3D slice blocks compose a complete full-resolution 3D reconstructed image. The composition process follows the following principles: (1) The first slice block takes the first $L \times 3/4$ slices; (2) The last slice block takes the last $L \times 3/4$ slices; (3) The rest of the slice blocks take the middle $L/2$ slices. Then all the selected slices are sequentially stitched into a complete 3D image with original size $F$.

The split-slice strategy can reduce the computational cost during training by a factor of $F/L$ (when $F = 320$ and $L = 16$, the computational cost can be reduced to just $1/20$ of the original). Furthermore, this method supervises all the $L$ slices in a block at each training iteration. But in the inference phase, only the middle $L/2$ slices are used for most of the slice blocks. These slices contain sufficient information from the surrounding slices and exhibit better reconstruction quality.

**3D-FasterFC based 3D sensitivity map estimation and global receptive field reconstruction**: Now we describe how 3D-FasterFC can be used for the reconstruction method in the split-slice strategy. The core is to use 3D-FasterFC to form an encoder-decoder network which we called 3D-FasterFC-U-Net (Fig. 7). As shown in Fig. 4(b), one 3D-FasterFC-U-Net is used for estimating 3D sensitivity map and cascaded 3D-FasterFC-U-Nets are used for reconstruction. The cascade times are denoted by $T$. As mentioned above, split-slice reconstruction takes in $\overline{\mathbf{x}}^{KNet}$ and splits it into $2 \times F/L - 1$ slice blocks (as shown in the up-right of Fig. 4). During the inference phase, all slice blocks are fed into the reconstruction network (i.e., 3D-FasterFC-U-Net) for refinement. Each block is first fed into a 3D FasterFC-U-Net to estimate the sensitivity maps. Different coil channels share one sensitivity map estimation network. For notation convenience, 3D-FasterFC-U-Net is also written as $FasterFCUNet_{3D}$. The sensitivity map estimation process can be formulated as:

$$\mathbf{S}_b = dSS \circ FasterFCUNet_{3D}(\overline{\mathbf{x}}_b^{KNet}), b = 1, ..., N_{\text{slice\_block}}. \quad (26)$$

Next, we fuse the multi-coil images using the estimated sensitivity maps:

$$\overline{\mathbf{x}}_b = \sum_{i=1}^{N} \mathbf{S}_{b,i}^* \mathbf{x}_{b,i}, b = 1, ..., N_{\text{slice\_block}}. \quad (27)$$

A multi-stage FasterFC-U-Net network with residuals takes $\overline{\mathbf{x}}_b$ as input for cascaded refinement:

$$\overline{\mathbf{x}}_b^{t+1} = FasterFCUNet_{3D}(\overline{\mathbf{x}}_b^t) + \overline{\mathbf{x}}_b^t, t = 1, ..., T. \quad (28)$$

Finally, all the $\overline{\mathbf{x}}_b^T$ merged into one full-resolution complex 3D image $\overline{\mathbf{x}}_c$ as depicted in Fig. 6. The output $\overline{\mathbf{x}}$ of FAS-Net then can be gotten after an absolution operation in the complex channel.

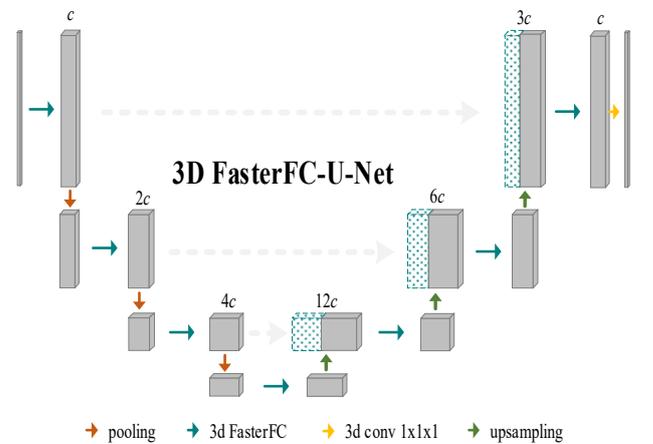

**Fig. 7.** The architecture of 3D FasterFC-U-Net.

## III. RESULTS

### A. Data

We demonstrate the effectiveness of FasterFC in multiple acquisition scenarios. For both 2D and 3D tasks, we employ Cartesian undersampling along the phase encoding direction. In the 2D single-coil and multi-coil tasks, we conduct experiments on the fastMRI dataset [31] using random Cartesian sampling with acceleration factor (AF) of 4 (where the middle 8% of the phase encoding lines are fully sampled and referred to as ACS region). This experimental setup is consistent with the requirements of the official fastMRI task.

In the 3D multi-coil study, we utilize the Stanford MRI Data dataset [32], which has more slice numbers. To assess the adaptability of our proposed FAS-Net and evaluate the generalization of FasterFC, we adopt equispaced Cartesian undersampling with an AF of 4 (i.e., 2-times-accelerated equispaced sampling in the two-phase encoding directions, and the middle 16% phase-encoded lines are fully sampled). It should be noted that because the Stanford MRI Data dataset employs an elliptical sampling strategy, the total acceleration factor is approximately 5. Schematic illustrations of the two undersampling trajectories mentioned above are provided in Fig. 8.

*1) FastMRI*

The fastMRI knee dataset contains raw *k*-space data and DICOM images, acquired by 3T and 1.5T MR devices. The dataset is organized into many volumes, each consisting of approximately 36 slices. The training, validation, and test sets comprise 973, 199, and 108 volumes, respectively. The data was acquired using a Cartesian 2D Turbo Spin Echo (TSE) sequence with a 15-channel phased array coil in the multi-coil dataset. Furthermore, the fastMRI dataset includes single-coil data generated from the multi-coil data. We verified the effectiveness of FasterFC on fastMRI single-coil dataset and demonstrated that FasterFC-U-Net has a faster calculation speed. Additionally, we validated the efficacy of FasterFC on the fastMRI multi-coil knee dataset.

We do not perform 3D experiments with the fastMRI dataset because it has few slices, and 3D reconstruction networks will not be able to take full advantage of the slice dimension.

*2) Stanford MRI Data*

We conducted 3D experiments on the Stanford MRI Data dataset, which was obtained from the Stanford Fully sampled 3D FSE Knees project. The data were collected using an 8-channel phased array coil by a 3T GE MEDICAL SYSTEMS scanner. The sequence parameters were defined as follows: matrix size was 320×320×1, field of view was 160*mm*×160*mm*×153.6*mm*, number of slices was 256, and trajectory was cartesian. The Stanford MRI Data dataset contains 19 volumes. We divided the 19 volumes into 14, 3, 2 for the training, validation, and test set. As mentioned in [44], Stanford MRI Data contains raw data collected at isotropic resolution. So, its data type is perfect for 3D reconstruction experiments. However, due to the small number of volumes, there are few 3D methods based on Stanford MRI Data that can achieve performance superior to 2D methods. To address this limitation, we propose FAS-Net, which employs single-to-group algorithm and FasterFC methods to improve performance. Additionally, we employed a split-slice training strategy to increase the number of training iterations while reducing the amount of computation. These modifications enabled FAS-Net to achieve superior reconstruction performance compared to the 2D state-of-the-art method, even when the amount of data is limited.

### B. Experimental Setup

*1) Metrics*

Averaged Structural Similarity (SSIM), Peak Signal to Noise Ratio (PSNR), and Normalized Mean Squared Error (NMSE) are adopted to measure the quality of reconstructed images [4] [31]. The parameters and definitions of the above three metrics are the same as described in the fastMRI paper [31].

*2) Implementation Details*

*a) 2D single-coil experimental implementation details*

For 2D single-coil experiments, we compare Faster-U-Net (Section III-B) with traditional U-Net baseline and FFC-U-Net. The experimental configuration of U-Net is consistent with the experiments in [31]. FFC-U-Net improves U-Net by changing each convolutional layer in the network to FFC [12]. Faster-U-Net replaces the network's adjacent two convolutional layers with one FasterFC module (as described in Fig. 2). Such comparative experiments can help understand the effectiveness of FasterFC, and can analyze various properties of the network, including performance, the number of parameters and calculation speed. The entry channel number *c* of these three networks is all 32, and downsampling rate is 4. The experiments are conducted on the fastMRI knee single-coil dataset.

*b) 2D multi-coil experimental implementation details*

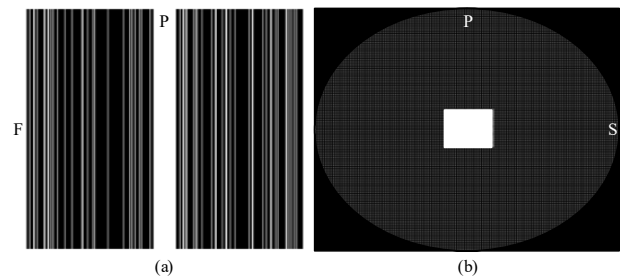

**Fig. 8.** Undersampling masks used in (a) 2D and (b) 3D reconstruction tasks.

TABLE I
RESULTS OF THE 2D SINGLE-COIL EXPERIMENT ON FASTMRI DATASET. BEST RESULTS ARE IN BOLD FONT

| Method | $c$ | $L$ | #param.(M) ↓ | GFLOPs ↓ | Training time (h) ↓ | Reconstruction time (ms) ↓ | NMSE ↓ | PSNR↑ | SSIM↑ |
|---|---|---|---|---|---|---|---|---|---|
| U-Net | 32 | 4 | 7.8 | 44.1 | **1.1** | **12.2** | 0.0379 | 31.43 | 0.7327 |
| FFC-U-Net | 32 | 4 | **6.6** | 37.2 | 4.3 | 49.0 | 0.0360 | 31.85 | 0.7390 |
| FasterFC-U-Net | 32 | 4 | 6.7 | **33.4** | 2.6 | 25.5 | **0.0359** | **31.87** | **0.7391** |

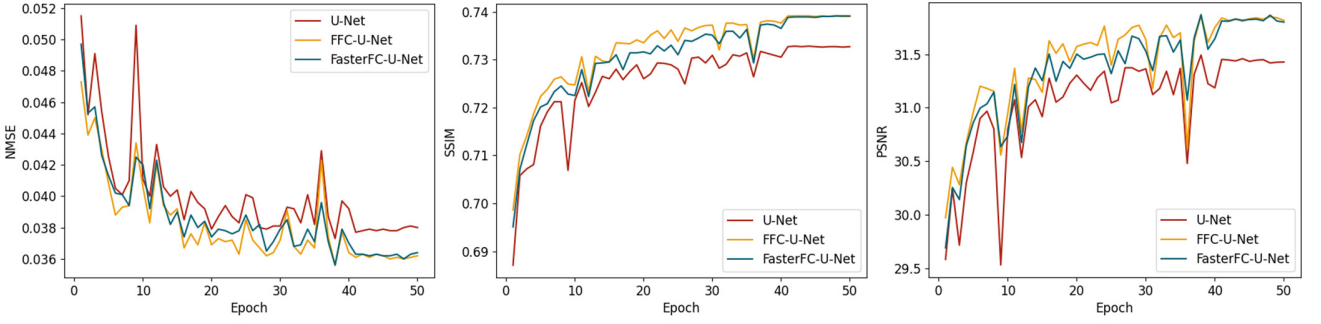

Fig. 9. Comparison of FasterFC-U-Net with image-domain U-Net and FFC-U-Net on validation set.

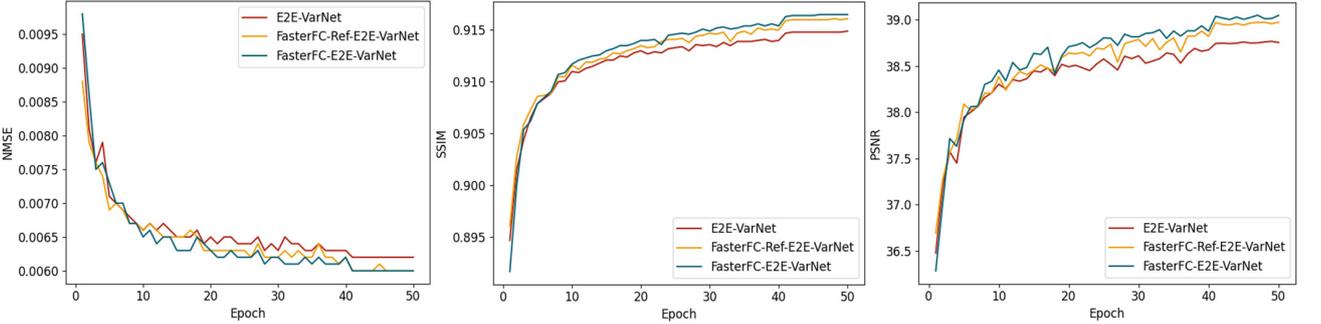

Fig. 10. Comparison of FasterFC-E2E-VarNet with E2E-VarNet and FasterFC-Ref-E2E-VarNet on validation set.

For the 2D multi-coil experiments, we verify the superiority of the proposed FasterFC-E2E-VarNet (Section Ⅲ-C) over classic E2E-VarNet. In our 2D multi-coil experiments, we developed two improved networks: (1) FasterFC-Ref-E2E-VarNet, which replaces U-Nets in the refinement modules of E2E-VarNet with FasterFC-U-Net, and (2) FasterFC-E2E-VarNet, which replaces U-Net used for estimating sensitivity maps with FasterFC-U-Net in FasterFC-Ref-E2E-VarNet. By comparing the results of these three experiments, we can analyze the ability of FasterFC to enhance the performance of sensitivity map estimation and reconstruction. All experimental configurations were kept the same as in [7], except for the number of cascades, which was set to 4 in our experiment.

*c) 3D multi-coil experimental implementation details*

The proposed FAS-Net (Section III-D) is compared with the following eight state-of-the-art 2D and 3D algorithms: CS [31], 2D U-Net Baseline [31], KIKI-net [8], D5C5 [6], ComplexMRI [33], Deep-SLR [30], E2E-VarNet [7], and 3D U-Net [45]. All experiments use the same mask matrix as FAS-Net, which is shown in Fig. 8(b). We used the same parameters in [31] for CS reconstruction. The 2D U-Net Baseline, a classic 2D spatial domain method, is implemented as [31]. KIKI-Net, D5C5, and ComplexMRI are three state-of-the-art methods utilizing the iterative network architecture. KIKI-net is also a dual-domain multi-stage reconstruction method to compare the effectiveness of FAS-Net's multi-stage strategy. ComplexMRI utilizes the complex convolution to process the complex values in the Stanford MRI data. Deep-SLR and E2E-VarNet are implemented strictly according to the original paper [7], [30] and source code. The 3D U-Net is a method that can flexibly adjust network parameters to adapt to the size of the memory of the experimental equipment, where the entry channels are set to 16, and the downsampling rate is set to 3. We specify that the metrics are computed on slices along the $F$ dimension of the volumes in the 3D case. This is convenient for us to use the same mask for different methods.

For the settings of 3D-RAKI in FAS-Net, the kernel sizes are set to $(5,2,2)$, $(1,1,1)$, $(3,2,2)$ for the three 3D convolutional layers, the kernel numbers of the first two layers are 32 and 8, the learning rate is 0.003, and the number of training epochs is 1000. For the details of 3D K-Net in group reconstruction, the entry channel number is 16, the downsampling rate is set to 3, the initial learning rate is 0.001 and will decrease to 0.0001 from 240 epochs, and the total number of training epochs is 250. For the settings of FasterFC-U-Net in split-slice training, the entry channel number and downsampling rate of network used for sensitivity map estimation are respectively 8 and 2, used for MR image refinement are set to 18 and 4. $T$ is set to 2, and $L$ is set to 16. The learning rate is also set to 0.001 for the first 160 epochs and will decrease to 0.0001 in the later 100 epochs. The PyTorch framework is adopted for model implementation. A machine with NVIDIA RTX A6000 GPUs is used.

*C. Quantitative Results*

*1) Results on the 2D Single-Coil FastMRI Dataset*

The quantitative results in Table I indicate that FasterFC-U-Net provides optimal balance between reconstruction performance and speed on the 2D single-coil fastMRI dataset. As depicted in Fig. 9, NSEM, PSNR, and SSIM curves are plotted against the training epoch number. Table I shows that FasterFC-U-Net outperforms other methods in terms of reconstruction metrics and GFLOPs while utilizing fewer parameters than U-Net. Although the NMSE, PSNR, and SSIM of FasterFC-U-Net is slightly better than those of FFC-U-Net, the reconstruction time in inference stage of FasterFC-U-Net is almost half of that of the FFC-U-Net. This is why we call our method FasterFC.

*2) Results on the 2D Multi-Coil FastMRI Dataset*

Table II demonstrates that FasterFC-E2E-VarNet outperforms the comparison methods on the fastMRI multi-coil dataset. Fig. 10 illustrates how reconstruction performance varies with the epoch number. By replacing all the U-Net in the refinement modules of E2E-VarNet with FasterFC-U-Net, FasterFC-Ref-E2E-VarNet achieves better NMSE, SSIM, and PSNR with fewer parameters and GFLOPs. This result highlights that the global receptive field of FasterFC aids in improving the network's reconstruction ability. Furthermore, FasterFC-E2E-VarNet improves the final reconstruction performance by replacing the U-Net for sensitivity map estimation with FasterFC-U-Net. This enhancement improves the quality of the sensitivity map estimation while also reducing the number of network parameters and GFLOPs.

TABLE II
RESULTS OF THE 2D MULTI-COIL EXPERIMENT ON FASTMRI DATASET. BEST RESULTS ARE IN BOLD FONT

| Method | #param. | GFLOPs | NMSE | PSNR | SSIM |
|---|---|---|---|---|---|
| E2E-VarNet | 10.3 M | 97.7 | 0.0062 | 38.75 | 0.9149 |
| FasterFC-Ref-E2E-VarNet | 9.0 M | 84.8 | 0.0060 | 38.97 | 0.9161 |
| FasterFC-E2E-VarNet | **8.9 M** | **75.6** | **0.0059** | **39.05** | **0.9165** |

*3) Results on the 3D Multi-Coil Stanford MRI Data Dataset*

Table III presents the quantitative results of the proposed FAS-Net and other state-of-the-art methods. The results demonstrate that FAS-Net outperforms all the other approaches. Among deep learning methods, 3D-U-Net, which has a small number of training iterations, achieves the worst results. By increasing the training iterations with the split-slice strategy and utilizing the single-to-group algorithm and FasterFC, FAS-Net achieves the best reconstruction performance.

TABLE III
RESULTS OF THE 3D MULTI-COIL EXPERIMENT ON STANFORD MRI DATA DATASET. BEST RESULTS ARE IN BOLD FONT

| Method | NMSE↓ | PSNR↑ | SSIM↑ |
|---|---|---|---|
| CS | 0.0498 | 35.19 | 0.7056 |
| 3D U-Net | 0.0204 | 39.00 | 0.9293 |
| 2D U-Net | 0.0155 | 40.22 | 0.9368 |
| KIKI-net | 0.0146 | 40.45 | 0.9390 |
| D5C5 | 0.0142 | 40.58 | 0.9395 |
| ComplexMRI | 0.0141 | 40.62 | 0.9399 |
| Deep-SLR | 0.0140 | 40.63 | 0.9402 |
| E2E-VarNet | 0.0140 | 40.64 | 0.9415 |
| FAS-Net | **0.0131** | **40.93** | **0.9447** |

*D. Qualitative Results of 3D Multi-Coil Stanford MRI Data Dataset*

The images displayed in Fig. 11 were randomly selected to confirm the quantitative results through qualitative assessment. Notably, the CS method yields extremely poor visual performance due to the sensitivity map estimation bias and underutilization of prior knowledge. Among deep learning methods, the 3D-U-Net performs the worst due to its smaller number of training iterations. Fig. 11 shows that FAS-Net outperforms its competitors in recovering the image's inner contrast and having a sharper structure.

IV. CONCLUSION

In this work, we proposed a novel convolutional operator called Faster Fourier Convolution (FasterFC) to replace the two consecutive convolution operations of CNNs commonly used in accelerated MR image reconstruction tasks. Table I demonstrates that FasterFC is roughly two time faster than FFC in 2D single-coil image reconstruction and the reconstruction accuracy of FasterFC based method (i.e., FasterFC-U-Net) is slightly but consistently better than the FFC-based method (i.e., FFC-U-Net). In addition, the 2D multi-coil experiments illustrated that using FasterFC in the sensitivity estimation network and the reconstruction network can effectively improve the reconstruction quality. Finally, based on the high efficiency of FasterFC, we proposed a novel 3D multi-coil MR reconstruction framework, FAS-Net. Cooperating with single-to-group algorithm, split-slice strategy, and multi-stage reconstruction, FAS-Net achieved the best reconstruction performance for 3D high-resolution $(320, 320, 256)$ multi-coil (eight) accelerated data than other state-of-the-art 2D and 3D methods.

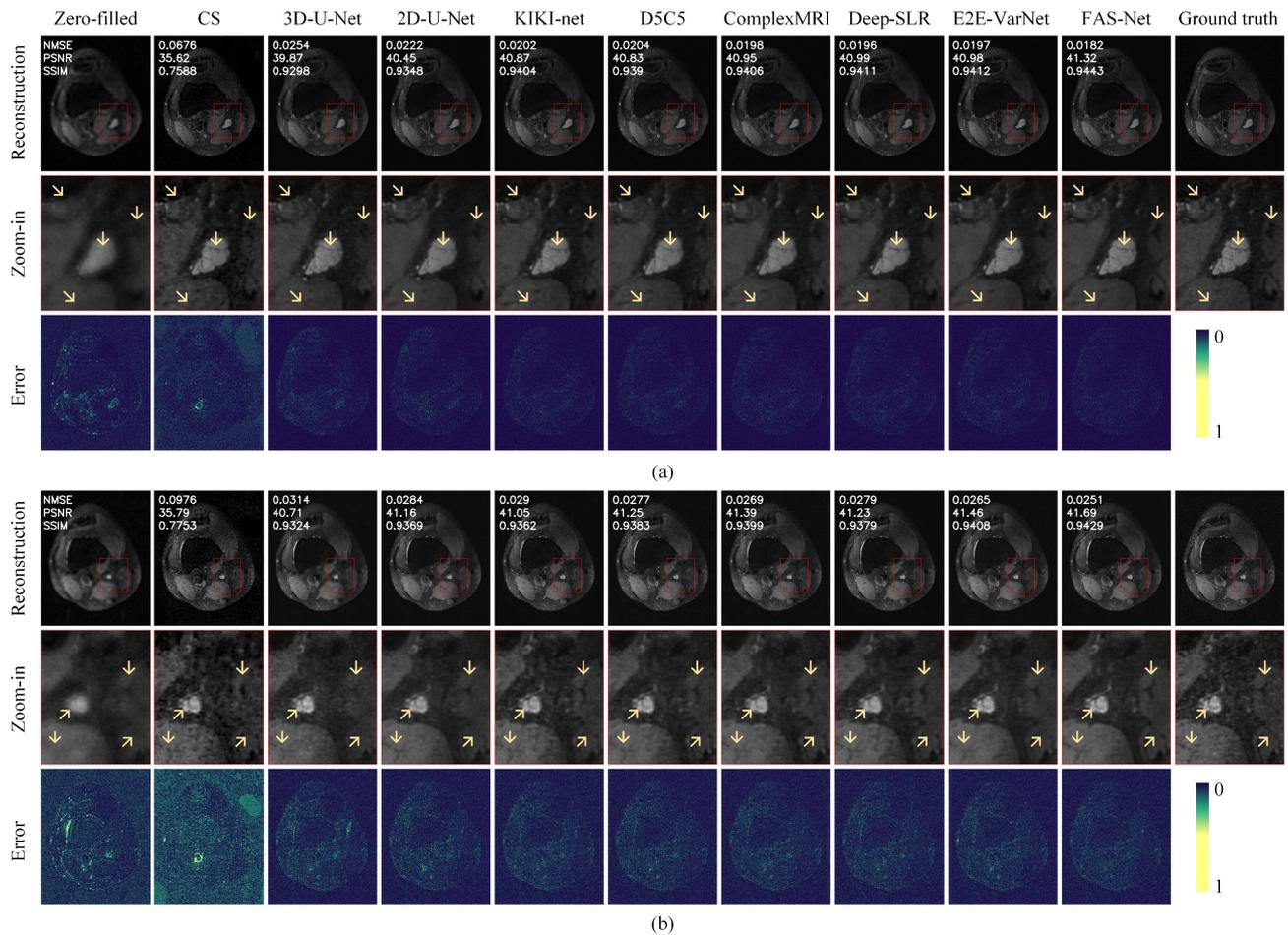

**Fig. 11.** Examples of reconstructed images and the ground truths in 3D multi-coil MRI data dataset. The yellow arrows in the zoomed cartilage region show minute details better preserved by the proposed method over other state-of-the-art methods.

The total reconstruction time of FAS-Net is about 82 seconds per 3D high-resolution multi-coil MR data. The E2E-VarNet method takes about 64 seconds ($0.2 \times 320$) to reconstruct all slices, slightly faster than FAS-Net. The bottleneck limiting the reconstruction speed of FAS-Net is that the calculation speed of 3D-RAKI (about 75s). The remaining two stages take 0.07 and 6.93 seconds, respectively, which can be ignored. In the future, we plan to design a faster single reconstruction method (i.e., Faster-RAKI) that can simultaneously calculate reconstruction networks for different coil in parallel, thereby increasing the reconstruction speed of FAS-Net by almost $N$ times.